\documentclass[reprint,pra,aps,superscriptaddress,floatfix,showpacs,tightenlines,twocolumn]{revtex4-1}
\usepackage{amsmath}
\usepackage{amssymb}
\usepackage{amsfonts}
\usepackage{dcolumn}
\usepackage{epsfig}
\usepackage{subfigure}
\usepackage[dvips]{color}
\usepackage{bm}
\usepackage{times}
\usepackage{amsthm}
\usepackage{graphicx}

\begin{document}
\title{Quantum Metrology via Repeated Quantum Nondemolition Measurements in ``Photon Box''}
\author{Yu-Ran Zhang}
\affiliation{Beijing National Laboratory for Condensed Matter Physics, Institute
of Physics, Chinese Academy of Sciences, Beijing 100190, China}
\author{Jie-Dong Yue}
\affiliation{Beijing National Laboratory for Condensed Matter Physics, Institute
of Physics, Chinese Academy of Sciences, Beijing 100190, China}

\author{Heng Fan}
\email{hfan@iphy.ac.cn}
\affiliation{Beijing National Laboratory for Condensed Matter Physics, Institute of Physics, Chinese Academy of Sciences, Beijing 100190, China}
\affiliation{Collaborative Innovation Center of Quantum Matter, Beijing 100190, China}
\date{\today}
\begin{abstract}
In quantum metrology schemes, one generally needs to prepare $m$ copies of $N$ entangled particles,
such as entangled photon states, and then they are detected in a  destructive process to
estimate an unknown parameter. Here, we present a novel experimental scheme for estimating this parameter by
using repeated indirect quantum nondemolition measurements in the setup called ``photon box''.
This interaction-based scheme is able to achieve the phase sensitivity scaling as
$1/N$  with a Fock state of $N$ photons.
Moreover, we only need to prepare one initial $N$-photon state and it can be used repetitively for
$m$ trials of measurements. This new scheme is shown to sustain
the quantum advantage for a much longer time than the damping time of Fock state and be more robust
than the common strategy with exotic entangled states.
To overcome the $2\pi/N$ periodic error in the estimation of the true parameter,
we can employ a cascaded strategy by adding a real-time feedback interferometric layout.
\end{abstract}


\maketitle

\section{Introduction}
Quantum parameter estimation, the emerging field of quantum technology, aims to use
entanglement and other quantum resources to yield higher statistical precision of a parameter
$\theta$  than purely classical approaches \cite{Lloyd2006,Lloyd2011}. The precision of
estimation of  $\theta$ will depend on the available resources used in the measurement. It has
been shown that standard quantum limit (SQL) or called shot noise limit scaling as $\delta\theta\simeq{1}/\sqrt{N_{\textrm{tot}}}$ with
$N_{\textrm{tot}}$ the number of particles can be surpassed by using coherent light with squeezed vacuum
\cite{Caves}. It is also commonly considered that using non-Gaussian states like NOON states \cite{NOON} and quantum
entanglement allows one to achieve a sub-shot noise accuracy. Heisenberg limit scaling as $\delta\theta\simeq1/N_{\textrm{tot}}$
is the ultimate limit set by quantum mechanics. Recently, some works \cite{zzb1,zzb2,zzb3}
have shown that, without prior information, sub-Heisenberg estimation strategies are
ineffective. There are also some papers showing that the Heisenberg limit can be saturated without
the use of any exotic quantum entangled states \cite{ent-free,interaction}.
Interferometric strategies with nonlinear phase encoding are investigated in Refs.~\cite{n1,n2}.  Practical quantum
metrology considering the impact of noise has been considered and studied in Refs.~\cite{noise1,noise2,noise3}.
The technique of quantum parameter estimation figures in several metrology platforms, including
optical interferometry \cite{opt1,opt2,opt3,IRS}, atomic systems \cite{atom1,atom2}, and Bose-Einstein
condensates \cite{BEC1,BEC2,BEC3,BEC4,BEC5,BEC6}. In addition, it is at the heart of many modern technologies and researches,
such as quantum clock synchronization \cite{clock1,clock2}, quantum imaging \cite{imaging}, and
gravitational wave observation \cite{gravity}.

General parameter estimation procedure can be divided into three distinct sections: probe
preparations, interaction between the probe and the system, and the probe readouts
\cite{Lloyd2011}. These three sections will be repeated many times before the final construction of
the estimation of $\theta$. Most of the quantum parameter estimation strategies require
preparation of $m$ copies of entangled states ($m$ is large enough). However, these states are extremely
difficult to generate and fragile to the impact of decoherence. Therefore, the method of quantum
nondemolition (QND) measurements \cite{RMP} initially with entanglement-free states may be a
suitable and practical way to overcome these challenges. The QND measurements dating back to
as early as the 1920s realize ideal projective measurements that leave the system in an
eigenstate of the measured observable \cite{QND}. With these ideal projective measurements
performed on an initial coherent state, Fock states and ``Schr\"{o}dinger cat'' states can be
prepared and reconstructed \cite{Haroche2008}. Moreover, with appropriate feedback loops,
it is possible to prepare on demand photon states and subsequently reverses the effects of
decoherence \cite{Haroche2011}. With these merits, we can foresee the widespread applications
of this techniques in quantum information and quantum metrology.

In this paper, we present a practical proposal for realizing quantum parameter estimation in
``photon box''  \cite{RMPH,Haroche1996,Haroche2007a,Haroche2007l} via QND  measurements.
We show that, with single-mode Fock state of $N$ photons in the ``photon box'', this proposal can
estimate the parameter $\theta$ within a scaling of $1/N$.
Unlike other quantum
metrology strategies, our proposal has this advantage-the state of  photons can be used circularly.
Thus, our scheme performs better than the strategy with NOON state when the total resource is
taken into consideration.
We also investigate our QND
metrology scheme with cavity damping.
It is shown that our scheme can sustain
the quantum advantage for a longer time than the damping time of Fock state and is more robust
than the interferometric strategy with exotic entangled states.
An improved cascaded estimation scheme is also proposed by adding a real-time feedback
interferometric layout \cite{Haroche2011}, with which the common $2\pi/N$ periodic error
can be handled.
The experimental feasibility of our
proposals can be justified with current laboratory parameters \cite{RMPH}.
We also discuss the possible applications of our QND metrology scheme.

\section{Parameter estimation via QND measurements in ``photon box''}

In our QND metrology proposal, the experimental setup is similar to the one discussed in
Refs.~\cite{Haroche1996,Haroche2007a,Haroche2007l,RMPH} and is shown in Fig.~1(a). The core
of this setup is a ``photon box'', which is an open high $Q$ cavity $\textrm{C}$ made up of two superconducting
mirrors facing  each other (Fabry-P\'{e}rot  configuration). QND probe atoms, generated form the
atomic resource $\textrm{S}$, are prepared in circular Rydberg states and travel along the transverse
direction of the cavity axis. The atoms cross the cavity $\textrm{C}$ sandwiched between two auxiliary
low $Q$ cavities $\textrm{R}_{1}$ and $\textrm{R}_{2}$ before being detected in the detector $\textrm{D}$.
The $\textrm{R}_{1}$-$\textrm{C}$-$\textrm{R}_{2}$ structure can be regarded as a Ramsey
interferometry. The microwave field stored in the cavity $\textrm{C}$ is with frequency $\omega_{\textrm{C}}/2\pi$.
The atomic frequency is $\omega/2\pi$ and is detuned from the cavity mode by $\delta/2\pi$
($\delta=\omega_{\textrm{C}}-\omega$).

\begin{figure}[t]
 \centering
\includegraphics[width=0.4\textwidth]{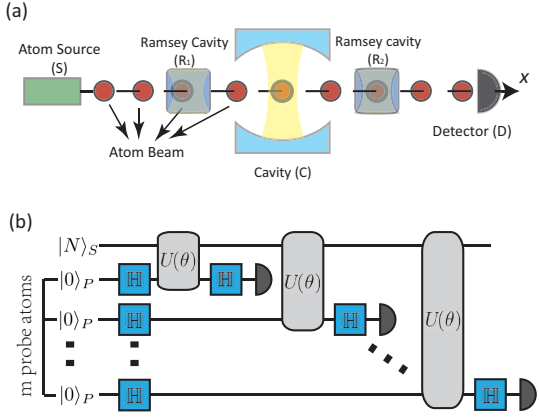}\\
\caption{(color online). {Setup and experimental sequence.}  (a) The cavity QED Ramsey interferometer for phase estimation.
The Rydberg atoms, prepared in state $|\uparrow_{z}\rangle$, are generated in the atom source $\textrm{S}$.
The interaction between the Rydberg atoms and microwave pulse in auxiliary cavities $\textrm{R}_{1}$
and $\textrm{R}_{2}$ perform Hadamard gate operation. The unknown parameter is imprinted by the
 interaction of the atom and the superconducting cavity $\textrm{C}$. After crossing the
 $\textrm{R}_{1}$-$\textrm{C}$-$\textrm{R}_{2}$ interferometric arrangement, the states of atoms are detected
 in the detector $\textrm{D}$. (b) Diagram for our sequential strategy. $m$ probe atoms are used. The
 Fock state of photons in $\textrm{C}$ stays unchanged after each QND measurement.}
\end{figure}

Suppose that the state of photons in the cavity $\textrm{C}$ is in a superposition of Fock states with
different photon numbers $|\psi\rangle_{S}=\sum_{n}c_{n}|n\rangle$. One Rydberg atom is prepared
in states $|\varphi\rangle_{P}=|\uparrow_{z}\rangle$; and afterwards, for simplicity,  we replace
$|\uparrow_{z}\rangle$ and $|\downarrow_{z}\rangle$ with $|0\rangle$ and $|1\rangle$. Both pulses
in Ramsey cavities $\textrm{R}_{1}$ and $\textrm{R}_{2}$ are acting as an Hadamard operation on each
atom which is written as $\mathbb{H}=(^{1}_{1}{}_{-1}^{\ \ 1})/\sqrt{2}$ and transforms $|0\rangle$
and $|1\rangle$ to $|+\rangle=(|0\rangle+|1\rangle)/\sqrt{2}$ and $|-\rangle=(|0\rangle-|1\rangle)/\sqrt{2}$,
respectively.
The interaction between the probe atom and photons contains an unknown parameter $\theta$
(see Appendix \ref{A} for details) and can be expressed as a unitary operator: \cite{Haroche2007a,PRA}
$\hat{U}_{SP}(\theta)=\exp\left[i{\theta}(\hat{n}_{S}+1/2)\hat{\sigma}^{z}_{P}/{2}\right]$ to the lowest order for small $\theta$,
where $\hat{n}_{S}$ is the photon number operator in the cavity $\textrm{C}$ and $\hat{\sigma}^{z}_{P}=(^{1}_{0}{}_{-1}^{\ \ 0})$
is the Pauli operator. The final state of photons and one probe atom after the probe atom passing through the
$\textrm{R}_{1}$-$\textrm{C}$-$\textrm{R}_{2}$ is expressed as
\begin{eqnarray}
|\Phi_{f}\rangle_{SP}=\mathbb{H}_{P}\hat{U}_{SP}(\theta)\mathbb{H}_{P}|\psi\rangle_{S}|\varphi\rangle_{P}.
\end{eqnarray}
We then perform the $\hat{\sigma}^{z}$ measurement on the atom in the detector, and the output
is $i=1$ or $-1$ with probability $p(i|\theta)=\sum_{n}c_{n}\cos^{2}[(n+1/2)\theta/2+(i-1)\pi/4]$.
Then, the photon state in $\textrm{C}$ is affected by this measure due to different outputs:
\begin{equation}
|\psi(i|\theta)\rangle_{S}=\sum_{n}\frac{c_{n}\cos[(n+1/2)\theta/2+(i-1)\pi/4]}{\sqrt{p(i|\theta)}}|n\rangle_{S}.
\end{equation}
It is easy to verify that $[\hat{\sigma}_{P}^{z},\mathbb{H}_{P}\hat{U}_{SP}(\theta)\mathbb{H}_{P}]|0\rangle_{P}=0$,
which is the general necessary and sufficient condition \cite{RMP} that the QND probe must satisfy.

Next, we present the experimental procedure for estimating a parameter, see Fig.~1(b). We consider that
the state of photons in the cavity is prepared as a  Fock state of $N$ photons, $c_{n}=\delta_{n,N}$.
Although generating a single mode Fock state of $N$ photons in the cavity $\textrm{C}$ is a challenging
task, it seems nowadays experimentally available \cite{Haroche2007a}. A general theoretic review of this
method is given in Appendix \ref{B}. It is also possible to prepare and
lock the field to on-demand photon number states by the real-time quantum feedback techniques
reported in Refs.~\cite{Haroche2011,HarochePRL}. Preparing the single-mode number
squeezing state is also helpful and urgent for ultrasensitive two-mode interferometry \cite{smerzi}.
It is easy to verify that, after the QND measurements procedure discussed above, a Fock state stays
unchanged \cite{JPA}.

Because $\langle\sigma_{z}\rangle=\cos[(N+1/2)\theta]$,  the parameter $\theta$
can be estimated from the readouts of $\sigma_{z}$ measurements performed on a sequence of probe atoms
interacting with the light in $\textrm{C}$.
For each probe, the probability for the readout $i=1$ or $-1$ is
$p(i|\theta)=\cos^{2}[(N+1/2)\theta/2+(i-1)\pi/4]$.
With the assumption that the estimation is
asymptotically unbiased, we can utilize Fisher information (FI)
$\mathcal{F}_{\theta}=\sum_{i}p(i|\theta)[\partial_{\theta}\ln p(i|\theta)]^{2}$
and Cram\'{e}r-Rao bound $\delta \theta={1}/{(\sqrt{m\mathcal{F}_{\theta}})}$ \cite{Lloyd2011}
to calculate the statistical precision of the estimation. FI is calculated as
$\mathcal{F}_{\theta}=(N+1/2)^{2}$ which leads to a lower bound:
\begin{eqnarray}
\delta \theta\geq\frac{1}{\sqrt{m}(N+1/2)}.\label{he}
\end{eqnarray}
Therefore, by using this QND metrology technique, our proposal is able to achieve the $1/N$
scaling accuracy of parameter estimate with only one initial $N$-photon Fock state for $m$ trials
of measurements.

The advantages of this QND metrology strategy are:  the initial photon state
is entanglement-free state which is more robust than the exotic states (e.g. NOON states), and
the Fock state stays unchanged after QND measurements and can be used repeatedly. Technically,
for our scheme, the total resource can be written as $N_{\textrm{tot}}=N+m\sim m$ for a  sufficiently large $m$, and the
lower bound is expressed as $\delta \theta\gtrsim1/(N\sqrt{N_{\textrm{tot}}})$. When $m$ copies of
NOON states are used to achieve the same accuracy, the total resource are
$N_{\textrm{tot}}=mN$ and the lower bound is
$\delta \theta^{\textrm{en}}\gtrsim1/\sqrt{NN_{\textrm{tot}}}$. Although both strategies do not achieve
the Heisenberg limit $1/N_{\textrm{tot}}$, given the
same total resources, our scheme gives $1/\sqrt{N}$ advantage compared with the strategy
using NOON states and $1/N$ advantage over SQL ${1}/\sqrt{N_{\textrm{tot}}}$.

\section{ Feasibility analysis in the real experiment}

Since the Bohr-Einstein photon box thought
experiment,  experiments with circular Rydberg atoms and Fabry-P\'{e}rot have become closest to
this goal. They have also led to fundamental tests of quantum theory and various demonstrations
of quantum information procedures \cite{RMPH}. Here, based on the developments and advances
made in the cavity quantum electrodynamics in the microwave domain, we discuss the feasibility
of our QND metrology scheme using current laboratory parameters.

It is reported in Ref.\ \cite{apll} an ultrahigh finesse Fabry-P\'{e}rot resonator
$\omega_{\textrm{C}}/2\pi=51.099$ GHz with cavity damping time $T_{\textrm{C}}=0.130\pm0.004$ s
and cavity quality factor $Q=4.174\times10^{10}$ at 0.8 K (mean number of blackbody photons $n_{b}=0.05$).
The damping rate is given as $\Gamma_{\textrm{C}}={1}/{T_{\textrm{C}}}={\omega_{\textrm{C}}}/{Q}$. In Ref. \cite{HarochePRL}, Rydberg
atoms are prepared by a pulsed process repeated at $\tau_{a}=82$ $\mu$s time intervals with
selected atomic velocity $v=250$ m/s. For $m$ trials of measurement, the total time is $t=m\tau_{a}$
and the photon-loss intensity in the cavity is written as $\eta(t)=1-\exp(-\Gamma_{\textrm{C}}t)$.
In fact, if we shorten the interval $\tau_{a}$, the number of measurement $m$
can be large with a low photon-loss intensity. This long damping time and QND detection technique
can stabilize the Fock state in the cavity $\textrm{C}$ and make our metrology scheme practicable.
In the next section, we will discuss the effect of cavity damping in detail.
The technologies of other experimental procedures, such as generation of Rydberg atoms and
polarizing measurement on the atoms, should be feasible and mature referring to Ref.~\cite{RMPH}.

\section{QND metrology with cavity damping}
The quantum metrological bounds in noisy systems
have become a focus of attentions because in real experiments there will always be some degree of
noise and limitation. The Fock state prepared in the cavity $\textrm{C}$ mainly suffers from cavity
damping. Given a certain the damping rate $\Gamma_{\textrm{C}}$, the interaction picture of reduced
density operator for the field in the cavity $\textrm{C}$  under the Born-Markov obeys the master
equation \cite{damping}:
\begin{eqnarray}
\dot{\rho}_{S}=&-&{\Gamma_{\textrm{C}}}n_{b}(\hat{a}\hat{a}^{\dag}\rho_{S}-2\hat{a}^{\dag}\rho_{S}\hat{a}+\rho_{S}aa^{\dag})/2\nonumber\\
&-&{\Gamma_{\textrm{C}}}(n_{b}+1)(\hat{a}^{\dag}\hat{a}\rho_{S}-2\hat{a}\rho_{S}\hat{a}^{\dag}+\rho_{S}\hat{a}^{\dag}\hat{a})/2,
\end{eqnarray}
where $\hat{a}$ ($\hat{a}^{\dag}$) is the annihilation (creation) operator for field in cavity.
Considering the radiation field with a reservoir at nearly zero temperature $n_{b}\ll1$, we
approximately express the density operator for the field by the well-known photon loss model:
$\rho_{S}=\sum_{k=0}^{N}(^{N}_{k})(1-\eta)^{k}\eta^{N-k}|k\rangle_{S}\langle k|$.

\begin{figure}[t]
  \centering
  \includegraphics[width=0.48\textwidth]{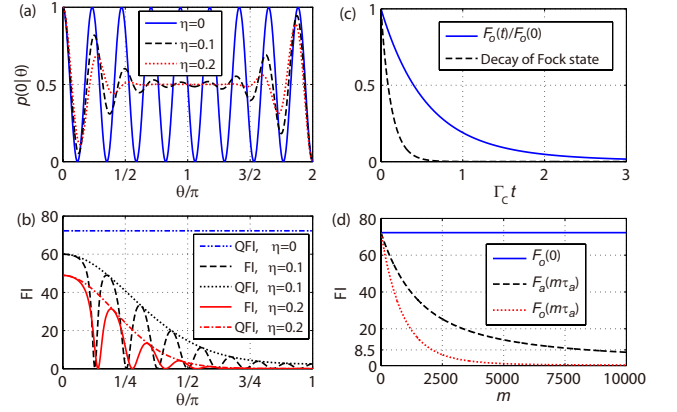}\\
  \caption{(color online). The photon number for the field in cavity $\textrm{C}$ is set as $N=8$ and cavity damping time is $T_{\textrm{C}}=0.130$ s.
  (a) Probability $p(0|\theta)$ against parameter $\theta$ for $\eta=0$, $0.1$ and $0.2$.  (b)
FI and QFI against parameter $\theta$ for $\eta=0$, $0.1$ and $0.2$.
(c) The decay of the optimal QFI $F_{o}$ compared with the decay of Fock state.
(d) The average QFI $F_{a}$ compared with the optimal QFI $F_{o}$ against number of measurements $m$ with measurement interval $\tau_{a}=82$ $\mu$s.}\label{f2}
\end{figure}

The initial probe state is still prepared as $\rho_{P}=|0\rangle_{P}\langle0|$.
The density operator form of the final state can be written as
$\rho_{SP}^{f}=\mathbb{H}_{P}U_{SP}\mathbb{H}_{P}\rho_{S}\otimes\rho_{P}\mathbb{H}_{P}U_{SP}^{\dag}\mathbb{H}_{P}$
and the final reduced state for probe atom is
\begin{eqnarray}
\rho_{P}^{f}=\frac{1}{2}\left(\begin{array}{c c}
1+r^{N}\cos(N\varphi) & -ir^{N}\sin{(N\varphi)}\\
ir^{N}\sin{(N\varphi)} & 1-r^{N}\cos(N\varphi)
\end{array}
\right)
\end{eqnarray}
where we set $r^{2}=1-4\eta(1-\eta)\sin^{2}\frac{\theta}{2}$ and
$\varphi=\frac{\theta}{2N}+\arctan\frac{(1-\eta)\sin\theta}{\eta+(1-\eta)\cos\theta}$.
Then we perform the $\hat{\sigma}^{z}$ measurement on the atom and obtain the
results $0$ and $1$ with probabilities $p(0,1|\theta)=[1\pm r^{N}\cos(N\varphi)]/{2}$.
Given photon number $N=8$, $p(0|\theta)$ is shown in Fig.~\ref{f2}(a) for different
values of lossy intensities. We can therefore calculate FI as
$\mathcal{F}_{\theta}={\{\partial_{\theta}[r^{N}\cos(N\varphi)]\}^{2}}/[{1-r^{2N}\cos^{2}(N\varphi)}]$
where we have used $\partial_{\theta}r=-{\eta(1-\eta)\sin\theta}/{r}$ and
$\partial_{\theta}\varphi=\frac{\eta(1-\eta)\cos\theta+(1-\eta)^{2}}{[\eta+(1-\eta)\cos\theta]^{2}+\sin^{2}\theta(1-\eta)^{2}}+\frac{1}{2N}$.
For $N=8$, FI is $\theta$ dependent for a nonzero $\eta$,
see Fig.~\ref{f2}(b).

By choosing the optimal measurement, we can obtain the maximum FI which is also
called quantum Fisher information (QFI). Given the spectral decomposition of final
reduced state for probe atom, $\rho_{P}^{f}=\sum_{i}p_{i}|i\rangle\langle i|$, QFI can
be written with condition $p_{i}+p_{j}\neq0$ as
\begin{equation}
\mathcal{F}_{Q}=2\sum_{ij}\frac{|\langle i|\partial_{\omega_{0}}\rho|j\rangle|^{2}}{p_{i}+p_{j}}
=N^{2}r^{2N}\left[\frac{|\partial_{\theta}(\ln r)|^{2}}{1-r^{2N}}+|\partial_{\theta}\varphi|^{2}\right]
\end{equation}
where we have used $p_{0,1}=({1\pm r^{N}})/{2}$ and
$|0,1\rangle=(|0\rangle\pm e^{iN\varphi}|1\rangle)/\sqrt{2}$.
Comparing FI and QFI in Fig.~\ref{f2}(b), we can achieve the optimal FI and QFI as we carefully
choose $\theta\rightarrow0$:
\begin{eqnarray}
F_{o}\equiv\lim_{\theta\rightarrow0}\mathcal{F}_{Q}=\left[(1-\eta)N+{1}/{2}\right]^{2}+\eta(1-\eta)N
\end{eqnarray}
where for large $N$, we obtain that $F_{o}\rightarrow [(1-\eta)N+1/2]^{2}$.
Therefore, with the two-step adaptive method based on Bayesian estimation \cite{referee}, the
optimal QFI can be achieved with the same measurement on the probe atoms used in the
noiseless case.

Although the lifetime of
the Fock state $|N\rangle$ is $1/(N\Gamma_{\textrm{C}})$, much less than the cavity damping time
$T_{\textrm{C}}$, the optimal QFI for this QND strategy decays much slower as shown in Fig.~\ref{f2}(c).
It has been recognized that photon losses in interferometers gradually blur the gain yielded by the special quantum
states for parameter estimation; even with the best strategy, asymptotically the improvement with respect
to standard light sources is not by a scale change but only by a limited constant factor \cite{noise1}. Unlike
the interferometric strategy with exotic states, e.g. NOON states, the QND metrology scheme with Fock
state will sustain the quantum advantage for a longer time than the damping time of Fock state. For instance,
given photon number $N=8$ and cavity damping time $T_{\textrm{C}}=0.130$ s, the quantum advantage
remains until $t\simeq0.172$ s for $F_{o}(t)\geq N+1/2$ and the number of trials can be $m\simeq2097$ for
time interval $\tau_{a}=82$ $\mu$s. We can also define the average QFI for $m$ trials of measurement as
$F_{a}=\sum_{i=0}^{m-1}F_{o}(i\tau_{a})/m$, with which we can write the lower bound of estimate accuracy for
decoherence  scenarios as $\delta_{dec}\theta\geq1/mF_{a}$.
Since we show in Fig.~\ref{f2}(d) that the quantum-enhanced estimation against decoherence  does not limit
the number of measurements to be too small ($m\lesssim 8444$), we can conclude that this QND metrology
scheme is expected to be robust against the cavity damping.

\section{QND metrology in cascaded scheme}
Let us assume that the phase to be estimated lies in the interval $\theta\in[-\pi,\pi)$. One common but
intractable problem in the quantum-enhanced metrology is the $2\pi/N$ periodic error in the estimation of the
true phase if $N\theta\notin[-\pi,\pi)$ \cite{2pie,clock2}. To address this problem we will next extend the
cascaded protocol reported in Ref.~\cite{cascaded} to our QND method. It is realizable with the help
of the mature technology of state control in the ``photon box'' \cite{RMPH}.

Our cascaded scheme employs $L$ successively larger Fock states of $2^{0},2^{1},\cdots,2^{L-1}$
photons. We use $m$ Rydberg atoms as the QND probe for each Fock state. The total resource
used in this cascaded scheme is $mN=m\sum_{j=0}^{L-1}2^{j}\simeq m2^{L}$. The interaction with
the Fock state consisting of $2^{j}$ photons picks up the phase $\Theta_{j}=2^{j}\theta\mod[-\pi,\pi)$,
where $j=0,\cdots,L-1$. The real phase to be estimated can be written in an exact binary representation
$\theta=2\pi\sum_{k=1}^{L-1}\frac{d_{k}}{2^{k}}-\pi+\frac{\Theta_{L-1}}{2^{L-1}}$,
with digits $d_{k}\in\{0,1\}$. By distinguishing whether the phase is shifted by $\pi$ or not, we can determine
the value of the bit $d_{k}$ according to the relation
$d_{k}=[2(\Theta_{k-1}+\pi)-(\Theta_{k}+\pi)]/2\pi$.
We should note that the rounding error \cite{HarochePRL} that occurs whenever $|\Theta^{est}_{j}-\Theta_{j}|>\pi/2$
can be neglected given a large number of trials $m$. The last group $(j=L-1)$ then yields a Heisenberg type
limited estimate of the parameter with accuracy
\begin{eqnarray}
\delta\theta_{\textrm{cas}}\geq\frac{1}{\sqrt{m}(2^{L-1}+1/2)}\simeq\frac{2}{\sqrt{m}(N+1/2)},
\end{eqnarray}
which is merely less sensitive by a constant $2$ compared with Eq.~(3).

\begin{figure}[t]
 \centering
\includegraphics[width=0.38\textwidth]{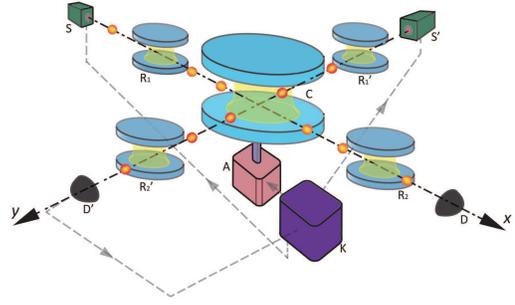}\\
\caption{(color online). {Layout of cascaded estimation scheme.} In addition to the QED Ramsey interferometer for phase
 estimation in $x$ axis, we need another real-time feedback interferometric setup to prepare the target
 Fock state in $y$ axis: $\textrm{R}_{1}^{'}$-$\textrm{C}$-$\textrm{R}_{2}^{'}$ interferometric arrangement
 and detector $\textrm{D}^{'}$. The detection results from detector $\textrm{D}'$ are sent to the computer
 based controller $\textrm{K}$. The controller $\textrm{K}$ analyzes each detection result and determines
 the the real translation amplitude $\alpha$ applied by actuator $A$.}\label{f3}
\end{figure}

To realize this cascaded QND metrology scheme, it is necessary to prepare and lock the field to
different photon number states during each QND measurement. This requirement can be fulfilled by
the real-time quantum feedback techniques reported in Refs.~\cite{Haroche2011,HarochePRL}. The
experimental layout of our proposal is shown in Fig.~\ref{f3}. The layout is supposed to work in two
modes: phase estimation mode and target state preparation mode. Since only one mode works at
the same time, we need two interferometric setups. In addition to the setup used for the phase
estimation mode, another interferometric setup is performed to prepare and stabilize the successively
larger target Fock states. The computer based controller $\textrm{K}$ controls the conversion between
those two modes (more details are shown in Appendix \ref{B}).

\section{Applications}
In our scheme, the unknown parameter expressed as
$\theta(v,z)=\sqrt{{2\pi}}{\Omega^{2}_{0}w}\cos^2({\omega_{\textrm{C}} z}/{c})/{v\delta}$ is determined
by the atom velocity $v$ and the position $z$ in the cavity $\textrm{C}$, see Appendix \ref{A} for
more details. Here,  we use the cylindrical coordinates $(R,z)$, $c$ is the speed of light in vacuum, $w$ is
the waist at center $(0,0)$ and $\Omega_{0}$  is the vacuum Rabi frequency at center. Therefore, this high
precision quantum-enhanced measurement can be used to detect and measure the mini-displacement of
the cavity $\textrm{C}$ along $z$ axis. On the practical perspective, a high sensitivity in $\theta$ leads to
the high sensitivity in the displacement $z$ when we measure the small displacement around the maximum
slope point $z=c\pi/(4\omega_{\textrm{C}})$. The accuracy can be obtained by straightforward error
propagation,
\begin{eqnarray}
\delta z=\frac{\delta \theta}{|d\theta/dz|}\geq\frac{z_{0}}{\sqrt{m}N}
\end{eqnarray}
where $z_{0}={\delta vc}/({\sqrt{2\pi}\Omega^{2}w\omega_{\textrm{C}}})$.
Using the current laboratory parameters shown in Sec.~III and
$\Omega_{0}/2\pi= 49$ kHz, $w=6$ mm and $\delta/2\pi=245$ kHz in Ref.~\cite{Haroche2011},
one obtains that $\delta z\gtrsim0.252/(\sqrt{m}N)$ mm. With $m=1000$ and $N=8$,
the sensitivity is $\delta z\gtrsim0.997$ $\mu$m and it can be improved by reducing the atom-cavity detuning and velocity of atom or increasing $N$ and $m$. However, these methods for improvement may seem challenging in
the real experiment, for instance, reducing the wavelength of the light in the cavity will then make it
difficult to place the atoms at the maximum slope point. Although this sensitivity by now seems several orders of magnitude worse than the
sensitivity needed for gravitational wave observation, our scheme will inspire future experiments
demonstrating quantum-enhanced metrology and may be helpful to prospective applications in other
experimental platforms.

\section{Discussions}
In this Letter, we have presented an experimental proposal for estimating an unknown parameter
in ``photon box'' by using the method of QND measurements. We have shown that initially with
Fock state of $N$ photons, the $1/N$ scaling accuracy of the estimation can be achieved.
Moreover, we do not need to prepare $m$ copies of initial state as other metrology schemes, which
will give a $1/\sqrt{N}$ advantage compared with the scheme using NOON states when the total resource is taken into consideration.
We also show that this sub-shot-noise estimation scheme is robust against cavity damping. The feasibility of our scheme
can be met by the current laboratory achievements and it can be improved via a cascaded scheme to
overcome the $2\pi/N$ periodic error. Furthermore,
this proposal with the help of QND measurements will also be an inspiration to other experimental
platforms \cite{NJP} for quantum metrology and quantum information techniques.  In addition,
our results should be of broad interest as many applications, such as clock synchronization and phase
imaging. 
Since generating and using  the NOON states with more than $3$ photons \cite{IRS} for quantum metrology is
still an arduous task, researchers have been able to generate Fock state with $7$ or even more photons in the
cavity. That is to say if our scheme can be realized in the experiments, it will be a great advance in
the research area of quantum metrology and quantum physics.
\begin{acknowledgements}
We would like to thank Augusto Smerzi and Mehdi Ahmadi for useful discussions.
This work was supported by the ``973'' Program (2010CB922904),
NSFC (11175248)
grants from the Chinese Academy of Sciences.
\end{acknowledgements}
\appendix
\section{Atom-Light Interaction in Cavity}\label{A}
We describe in this section the interaction between the Rydberg atom and photons in the cavity
in a concise form. This simple case will provide us the phase shift  linearly given by per photon.

The interaction can be described via the Hamiltonian of Jaynes-Cummings model \cite{BOOKl}
\begin{eqnarray}
\frac{\hat{H}}{\hbar}=\frac{\omega\hat{\sigma}^{z}_{P}}{2}+\omega_{\textrm{C}}\left(\hat{a}^{\dag}\hat{a}+\frac{1}{2}\right)+ g(\hat{a}\hat{\sigma}^{+}_{P}+\hat{a}^{\dag}\hat{\sigma}^{-}_{P})
\end{eqnarray}
where $\hat{a}$ ($\hat{a}^{\dag}$) is the photon annihilation (creation) operator in
the cavity $\textrm{C}$. Let the cavity $\textrm{C}$ contain $n$ photons. For large detuning
frequency, the atom-field states $|0,n\rangle$ and $|1,n\rangle$ evolve into dressed states
and are shifted in angular frequency units by \cite{prl2l}
\begin{eqnarray}
\Delta(\bm{r},n)\simeq\hbar(n+1/2)\Omega^2(\bm{r})/\delta.
\end{eqnarray}
where $\Omega(\bm{r})=\Omega_{0}\exp(-R^{2}/w^2)\cos(\omega_{\textrm{C}}z/c)$ is the
vacuum Rabi frequency  following the Gaussian distribution at cavity center $z=0$; here we use
cylindrical coordinates $(R,z)$ where $z$ is the position of the atom along the beam axis.
$c$ is the speed of light in vacuum, $w$ is the waist at center $(0,0)$ and
$\Omega_{0}$  is the vacuum Rabi frequency at center. The
difference of the phase imprinted on atomic states should be expressed as $(n+1/2)\theta(v,z)$
where $\theta(v,z)=\sqrt{{2\pi}}{\Omega^{2}(0)w}\cos^2({\omega_{\textrm{C}} z}/{c})/{v\delta}$ with
$w$ the waist at center and $v$ the atom velocity. Therefore, the interaction between
the probe atom and photons can be expressed with the parameter $\theta(v,z)$ as a unitary operator:
$\hat{U}_{SP}(\theta)=\exp\left[i{\theta}(\hat{n}_{S}+1/2)\hat{\sigma}^{z}_{P}/{2}\right]$ to the lowest order
for large detuning frequency. The value of the parameter $\theta(v,z)$ is determined by the
atom velocity and the position in the cavity $\textrm{C}$, which will be of great value in scientific
and engineering applications.

\section{Preparation of Fock States and Cascaded Scheme}\label{B}
In this section, we review two different  methods of preparation of Fock state in the ``photon box''.
The stochastic method is expected to require less equipments and be suitable for a demonstrative
experiment of quantum-enhanced metrology. The deterministic method needs additional experimental
devices and is able to generate a Fock state with on-demand photon number, which is helpful to the
cascaded strategy.
\subsection{Stochastic Approach}
\begin{figure*}[t]
 \centering
\includegraphics[width=0.92\textwidth]{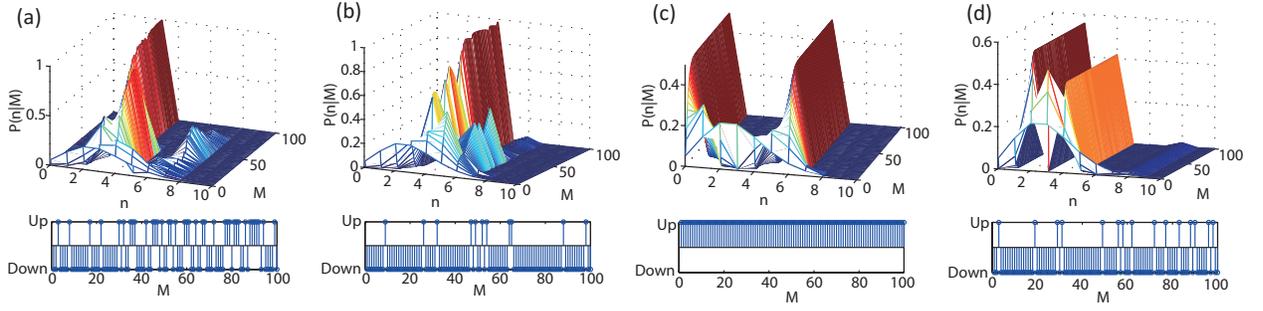}\\
\caption{(color online). {Numerical simulation of the indirected measurement procedures.} For each simulation,
the convergence event is obtained via Monte Carlo method and considers totally 100 atoms.
The state of light in the cavity is initially chosen
as a coherent  state $|\sqrt{3}\rangle_{S}$. $M$ is the number of atoms interacting
with the light in the cavity, $n$ represents the photon number and  $P(n|M)$ denotes
the photon number probability distribution after the $M$th atom flies through the cavity. The
2D diagram shows the detection result of $M$ atom in the sequence. Two
parameters are considered: $\theta_{s}^{(1)}=0.6$ for (a) and (b); $\theta_{s}^{(2)}=\pi/3$ for (c) and
(d).}\label{sf1}
\end{figure*}
We start with  $|\psi\rangle_{S}=\sum_{n}c_{n}|n\rangle$ and set the interaction parameter
$\theta_{s}$  at an appropriate and definite value such that $p(i|n)\neq p(i|n')$ for all possible
photon number $n\neq n'$, where $p(i|n)=\cos^{2}[(n+1/2)\theta_{s}/2-(i-1)\pi/4]$.  Suppose that
$M$ atoms cross the $\textrm{R}_{1}$-$\textrm{C}$-$\textrm{R}_{2}$
interferometric layout and are measured with operator $\sigma^{z}$. The sequence of measurement
results for $M$ probe atoms, called a event, is expressed as $\bm{\omega}_{M}=(i_{1},\cdots,i_{M})$,
where $i_{\mu}=1,-1$ and $\mu=1,2,\cdots,M$. The photon number distribution of the final state in
$\textrm{C}$ can be calculated as
\begin{equation}
P(n|\bm{\omega}_{M})=\frac{|c_{n}\cos^{\eta}[(n+\frac{1}{2})\theta_{s}/2]\sin^{\xi}[(n+\frac{1}{2})\theta_{s}/2]|^{2}}{Z(\bm{\omega}_{M})},
\end{equation}
where $Z(\bm{\omega}_{M})=\sum_{n}|c_{n}\cos^{\eta}[\frac{(n+{1}/{2})\theta_{s}}{2}]\sin^{\xi}[\frac{(n+{1}/{2})\theta_{s}}{2}]|^{2}$;
$\eta$ and $\xi$ are the number of $1$ and $-1$ in the event $\bm{\omega}_{M}$, respectively.
It has been proved in Ref.~\cite{PRA} that ($i$) this photon number distribution converges as
$M$ becomes infinity: $\lim_{M\rightarrow\infty}P(n|\bm{\omega}_{M})=\delta_{n,N}$, ($ii$) the probability for the state
in cavity converges to a Fock state $|N\rangle$ is $|c_{N}|^{2}$, and ($iii$) the convergence for
$\delta_{n,N}$ is exponentially fast.
Therefore, we can obtain a single mode Fock state $|N\rangle$ by this approach with probability
$|c_{N}|^{2}$ when $M$ is large enough. The final photon number $N$ can be determined via analyzing
the spin measurement results of the probe atoms $\bm{\omega}_{M}$. After
generating a Fock state of a nonzero and known photon number, we can perform the parameter
estimation without adjusting the experimental apparatus.

\begin{figure}[b]
\centering
\includegraphics[width=0.35\textwidth]{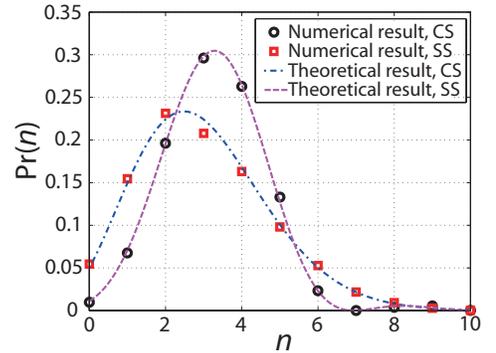}\\
\caption{(color online). {Reconstructed photon number distribution for coherent state and squeezed state.}
Photon number distributions for coherent state (CS) $|\sqrt{3}\rangle_{S}$ and
squeezed state (SS) $|\sqrt{3},0.5\rangle_{S}$ are plotted by dashed blue line and dash-dotted
magenta line, respectively.
$4000$ convergence events are simulated to
figure out the photon number distribution probability numerically.
}\label{sf2}
\end{figure}

Most commonly the initial state of the light in cavity $\textrm{C}$ is a coherent state
$|\psi\rangle_{S}=|\alpha\rangle_{S}$. The QND measurement procedures are numerically
simulated via Monte Carlo method and plotted in Fig.~\ref{sf1}. We observe the converging events of different
photon numbers in Fig.~\ref{sf1}(a) and \ref{sf1}(b). In Fig.~\ref{sf1}(c) and \ref{sf1}(d), we
present the situation that the special condition $q(i|n)=q(i|n')$ is saturated and the convergent states are
superposed Fock states. We also numerically
simulate the probability for the coherent state converging to a Fock state,
in Fig.~\ref{sf2}, which conforms with the experimental results in Ref.~\cite{Haroche2007a}.

The average Fisher  information for all possible Fock states from the initial
coherent state $|\alpha\rangle$ can be written as
\begin{eqnarray}
F_{\alpha}=\sum_{i=0}^{\infty}e^{-\bar{n}}\frac{\bar{n}^{i}}{i!}\left(i+\frac{1}{2}\right)^{2}=\left(\bar{n}+\frac{1}{2}\right)^{2}+\bar{n}
\end{eqnarray}
which leads to the Heisenberg-type lower bound: $\delta_{coh}\theta\geq1/[\left(\bar{n}+{1}/{2}\right)^{2}+\bar{n}]$.
Instead of the coherent state, an efficient method for improving this strategy is to use the squeezed state $|\alpha,\zeta\rangle_{S}$
which may be generated by first acting with the squeeze operator $\hat{S}(\zeta)$ 
on the vacuum followed by the displacement operator $\hat{D}(\alpha)$ \cite{squeeze}.
We can obtain via squeezed state a higher success rate for generating a useful Fock state
 for sub-shot-noise metrology due to its super-poissonian and narrower photon number statistics, see Fig.~\ref{sf2}.

\subsection{Deterministic Approach}
In order to prepare and lock the field to different photon number states during each QND measurement,
it is necessary to use real-time quantum feedback techniques reported in
Refs.~\cite{Haroche2011,HarochePRL} to fulfill this requirement. The experimental layout of our proposal is shown
in Fig.~3 in the main text. In addition to the estimation interferometric setup ($x$ direction) shown
in Fig.~3,  a quantum feedback setup is put in the $y$ direction: another atom source $\textrm{S}'$
generates test Rydberg atoms, and two auxiliary cavities $\textrm{R}_{1}^{'}$ and $\textrm{R}_{2}^{'}$ act
as two Hadamard gates. The information on the interaction between the test atom and the field is assumed
to be known. The measurement results of the test Rydberg atoms obtained by detector $\textrm{D}'$ are
sent to the computer based controller $\textrm{K}$. By analyzing each detection result, the controller
$\textrm{K}$ updates the photon distribution $p(n)=|c_{n}|^{2}$ and controls actuator $\textrm{A}$ to
feed cavity $\textrm{C}$ by diffraction on the mirror edges \cite{Haroche2011}. The controller $\textrm{K}$
analyzes each detection result to determine the real translation amplitude $\alpha$ applied by actuator
$\textrm{A}$ which minimizes the distance $d(\hat{\rho}_{t},\hat{D}(\alpha)\hat{\rho} \hat{D}^{\dag}(\alpha))$
\cite{Haroche2011} between the target $\hat{\rho}_{t}=|n_{t}\rangle\langle n_{t}|$ 
and the field estimation $\hat{\rho}$. Here $\hat{D}(\alpha)$ is the displacement operator. When the
controller $\textrm{K}$ finds that $\alpha\rightarrow0$, it stops atom source $\textrm{S}'$ and the target
Fock state has been prepared. It is reported that Fock states with photon numbers $n_{t}$ up to 7 can be
prepared with number photon number distribution peaked $p(n)=0.8\sim0.9$ \cite{HarochePRL}.
Then, controller  $\textrm{K}$ activates another atom source $\textrm{S}$ and
$m$ probe states are sent and detected to estimate the unknown phase. We will next show that this technique
based on  real-time quantum feedback is also necessary to realize the cascaded scheme \cite{cascaded}.
\begin{figure}[!t]
  \centering
  \includegraphics[width=0.35\textwidth]{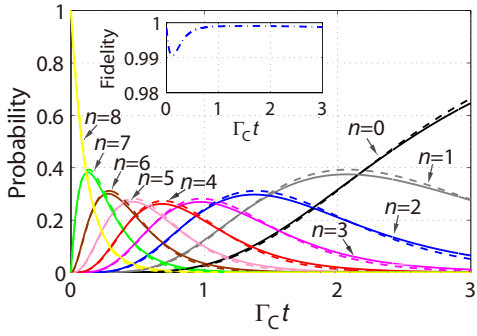}\\
  \caption{(color online). Probabilities of $\rho_{S}^{\textrm{exact}}$ and
$\rho_{S}$ against time for photon number $N=8$. Solid lines are for
$\rho_{S}^{\textrm{exact}}$ and dashed lines are for $\rho_{S}$. Inset:
Fidelity of $\rho_{S}^{\textrm{exact}}$ and
$\rho_{S}$ against time.}\label{bb}
\end{figure}

\subsection{QND Metrology in Cascaded Scheme}
One common but intractable problem in the quantum-enhanced metrology
is the $2\pi/N$ periodic error \cite{2pie} in the estimation of the true phase if $N\theta\notin[-\pi,\pi)$. This problem can
not be settled by simply adding an ancillary phase so that the cascaded protocol is proposed to solve it.
To realize the cascaded QND metrology scheme, it is necessary to prepare and lock the field to
different photon number states during each QND measurement.

The first target state starts with $n_{t}=2^{0}$ for $\hat{\rho}_{t}=|n_{t}^{j}\rangle\langle n_{t}^{j}|$.
When the target Fock state $n_{t}^{j}=2^{j}$ for $j=0,\cdots,L-1$ has been prepared, controller  $\textrm{K}$
activates another atom source $\textrm{S}$ and $m$ probe states are sent and detected to estimate
the unknown phase with value $\Theta_{j}$. After the estimation using this Fock state $|n_{t}^{j}\rangle$ is
finished, controller $\textrm{K}$ activates the feedback setup and the target state is changed to the
next Fock state $|n_{t}^{j+1}\rangle$. The last target Fock state that we consider has $n_{t}^{\max}=2^{L-1}$
photons. When all $\Theta_{j}$ for $j=0,\cdots,L-1$ have been obtained, we can calculate digits $d_{k}$
for $k=1,\cdots,L-1$ and retrieve the true value of phase $\theta$ within a accuracy presented in Eq.~(8).

As a brief summary, the layout is supposed to work in two modes: target state preparation mode and
phase estimation mode. Since only one mode works at the same time, we need two interferometric setups.
In addition to the setup used for the phase estimation mode, another interferometric setup is performed
to prepare and stabilize the successively larger target Fock states. The computer based controller
$\textrm{K}$ controls the conversion between those two modes.

\section{Approximation in QND Metrology with Cavity Damping}\label{C}
The exact solution of the master equation (4) in the main text can be expressed by the density operator:
$\rho^{\textrm{exact}}_{S}(t)=\sum_{n}q_{n}(t)|n\rangle_{S}\langle n|$ with probabilities \cite{1990}
\begin{equation}
q_{n}\simeq\frac{\sum_{k=0}^{n}\Big({}^{N}_{k}\Big)\Big({}^{n}_{k}\Big)n_{b}^{n-k}(1+n_{b})^{N-k}\eta^{k}(1-\eta)^{N+n-2k}}{[1+n_{b}(1-\eta)]^{N+n+1}}.
\end{equation}
In recent experiments \cite{RMPH}, the mean number of blackbody photons $n_{b}\simeq0.05\ll1$ at 0.8 K. Thus, we
simply choose $n_{b}=0$ with $N\leq n$ and use an approximate solution
$\rho_{S}=\sum_{k=0}^{N}(^{N}_{k})(1-\eta)^{k}\eta^{N-k}|k\rangle_{S}\langle k|$ to calculate the noisy case
with cavity damping. In Fig.~\ref{bb}, we compare the probabilities of these two states against time for photon
number $N=8$; solid lines are for $\rho_{S}^{\textrm{exact}}$ and dashed lines are for $\rho_{S}$. We also
show in the inset figure of Fig.~\ref{bb} that the fidelity of these two states $\rho_{S}^{\textrm{exact}}$ and
$\rho_{S}$ against time is always larger than 0.99. Therefore, this approximation with which we can obtain a
analytical result of Fisher information and quantum Fisher information is proper.

\end{document}